# Symmetry Theory of the Flexomagnetoelectric Interaction in the Magnetic Vortices and Skyrmions

B. M. Tanygin[a]

[a] Radiophysics Department, Taras Shevchenko Kyiv National University, 4G, Acad. Glushkov Ave., Kyiv, Ukraine, UA-03127

*Corresponding author:* B.M. Tanygin, Radiophysics Department, Taras Shevchenko Kyiv National University, 4G, Acad. Glushkov Ave., Kyiv, Ukraine, UA-03127.

*E-mail*: b.m.tanygin@gmail.com

*Phone*: +380-68-394-05-52

**Abstract.**

Symmetry classification of the magnetic vortices and skyrmions has been suggested. Relation between symmetry based predictions and direct calculation has been shown. It was shown, that electric dipole moment of the vortex is located inside the small vortex core. The antivortices and antiskyrmions do not carry the total core electric dipole induced by the flexomagnetoelectric interaction in the hexoctahedral cubic crystal. The volumetric bound electric charge is distributed around the core. Switching of the core electric dipole direction produces the switching of the core magnetization or vortex chirality and vice versa. The vortices and skyrmions with time-invariant enantiomorphism have two degenerative states: clockwise and counterclockwise state.

## 1. Introduction

Coupling mechanism between magnetic and electric subsystem in magnetoelectric materials [1] is of considerable interest to fundamentals of condensed matter physics and to the applications in the novel multifunctional devices [2] including low-power-consumption spintronic and magnonic devices. The magnetoelectric coupling is possible between homogeneous magnetization and electric polarization [3-7], and, also, between inhomogeneous magnetization and polarization [8-14]. Latter type of coupling was called as flexomagnetoelectric (FME) interactions [15-17]. The FME coupling is possible not only between existing magnetic and ferroelectric orderings (i.e. in the multiferroic material). It, also, describes inducing of the polarization in the region of the magnetic inhomogeneity [14]. The phenomenological theory of the FME coupling in any symmetry crystals of the cubic, tetragonal and orthorhombic families has been recently developed [18].

FME interaction has been investigated experimentally and theoretically in the micromagnetic structures like spin cycloids [12,19], magnetic domain walls (DWs) [8,9,13,16-17,20], magnetic (anti)vortices [14,15] and the vertical Bloch lines (theory [21] and preliminary experimental report [22]). The FME features are expected in case of the (anti)skyrmions [23] due to their "vortex" nature. The micromagnetic interaction which produces skyrmions usually is the antisymmetric Dzyaloshinskii-Moriya interaction with energy terms containing Lifshitz invariants [23]. The direct analogy with the FME properties and this interaction exists [18]. Consequently, skyrmion can be produced in the same way as the magnetic vortex.

The possible write/read process in the vortex random access memory or antivortex random access memory was suggested in Ref. [24]. The same practical applications are possible if case of the 0° (360°) domain wall [25] owing to the development of the magnetic memory elements using these domain walls [26]. The skyrmions and 0° domain walls are soliton solutions and can appear in the same medium [27]. The existence of skyrmions at room temperature improves the practicality of utilizing their potential for use in novel computer memories [28]. Skyrmions are able to remain stable against collapse to the atomic scale [29] which can give possibility to increase memory storage values. Skyrmions can give a new approach to electronic memory [28]. The magnetic dots that stabilize the magnetic vortex state are considered as elements of recording media and microwave signal-processing devices. The FME method of control of all these structures is of interest.

The magnetoelectric effects are closely related to the magnetic symmetry. The group-theoretical description of the FME coupling in all possible magnetic domain walls (including 0° domain wall) [30] and Bloch lines [21] was already developed. Building of the point symmetry and enantiomorphism classification of magnetic (anti)vortices and (anti)skyrmions with FME coupling is the purpose of the present work.

## 2. Symmetry classification of the magnetic vortices and skyrmions

Micromagnetic structures which are usually called as the magnetic (anti)vortex and magnetic (anti)skyrmion are localized 0D (e.g., in the nanodot) or 1D (e.g., in the nanowire) structures with different experimental conditions of the occurrence. Here, the dimensionality definition [31] was used. In order to develop complete symmetry classification of these structures it is necessary to give their formal definition. Skyrmion contains a small normal (relatively the sample plane) magnetization area surrounded by a large opposite area [23,27]. Internal structure of the skyrmion always contains rotation of the magnetization in the 3D space [23,27]. In contrast to skyrmion, the vortex structure contains the in-plane magnetization (rotation in 2D space) surrounded by domain with in-plane magnetization. An exception is made for the small vortex core [24,32] with out-of-plane magnetization which minimizes exchange energy in the center. Consequently, vortex structures have value $\theta = \pi/2$ with transition to the $\theta = 0$ or $\pi$ in core depending on the core polarization [24], where $\theta$ is an angle of the magnetization deflection from the normal (axis $Z$) of the sample plane (plane $XY$). Skyrmion structures always have continuous transition $0 \leq \theta \leq \pi$. It is necessary to compare experimental conditions of the occurrence of vortices and skyrmions.

Vortex structures are formed in soft magnetic thin-film with a lateral size of a few hundred nanometers to some microns when the demagnetization energy forces the magnetization to align parallel to the sample's surface. The vortices appear in the permalloy squares and disks [33]. The antivortices appear in the infinity-shaped, clover-shaped, and cross-like structures [34]. The vortex and antivortex states can be switched independently on the shape using FME coupling as it was proposed for the nanodots in the Ref. [15]. The electrode utilized for such switching can be cantilever tip of atomic force microscope [15].

In the non-centrosymmetric magnets the magnetic energy density contains inhomogeneous Dzyaloshinskii-Moriya exchange term [23,27] owing the skyrmion appearing (experimental observation in the thin layers of $(Co, Fe)Si$ [35]) which is not caused by the FME properties. The FME coupling produces the similar chiral symmetry breaking of the centrosymmetric magnets [36]. The periodic structure of skyrmions and antiskyrmions was investigated in Ref. [37]. The skyrmions was observed in the quantum Hall experiments [38].

The in-plane magnetization usually is determined by the following planar sinusoidal ansatz for the vortices [14] and skyrmions [23,27]:

$$\mathbf{M}_\perp(\rho, \varphi) = M_s\big(\mathbf{e_x} \cos(n\varphi + c\pi/2) + \mathbf{e_y} \sin(n\varphi + c\pi/2)\big), \qquad (1)$$

where $M_s$ is a saturation magnetization and $n$ is a winding number ($n > 0$ for vortices/skyrmions and $n < 0$ for antivortices/antiskyrmions). The structural configuration parameter $c$ is determined in the ranges $-2 \leq c \leq 2$. Let us start from the symmetry investigations of ansatz (1) (magnetic point group $G_V$) considering it as the axial time-odd vectors distribution.

The value $n = 1$ corresponds to the single vortex structures (Fig. 1a-c). All these structures have symmetry element $\infty_z$ and $m'_z$, which construct the limit Curie group. If $c = \pm 1$ then point group contains mirrors $m_\perp$, which are perpendicular to the $(XY)$ plane (Fig. 1c). If $c = \pm 2$ or $c = 0$ then the same mirrors are supplemented by the time reversal operation ($m'_\perp$) (Fig. 1a). If parameter $c$ is not integer then no mirrors containing axis $Z$ are present in the point group $G_V$ (Fig. 1b).

The range $n \in (-\infty, -1] \cup [2, +\infty)$ corresponds to structures without the axial symmetry $\infty_z$ (Fig. 1d-f). In this case, the group $G_V$ always contains mirror $m'_z$ and rotation axis $r'$, where order is given by the following expression:

$$r = 2|n - 1| \qquad (2)$$

Group $G_V$ contains $|n - 1|$ mirrors $m_\perp$ and the same quantity of the mirrors $m'_\perp$. Variation of the parameter $c$ at $n \neq 1$ leads to the rotation of ansatz around the axis $Z$. Symmetry classification of $G_V$ is provided using these basic symmetry transformations (table 1). The limit magnetic symmetry theory [39] was used. The enantiomorphism classification [40] was used. The vortices and skyrmions with the time-invariant enantiomorphism have two degenerative states: clockwise and counterclockwise state. The symmetries of the plane vortex around the core $G_V$ and vortex with core or skyrmion $G_C$ are separated.

**Table 1.** Symmetry classification of the magnetic vortices and skyrmions

| Winding number $n$ | $c$ | Point group $G_V$ (vortex around the core)* | Point group $G_C$ (skyrmion, vortex with core) | Enantiomorphism of $G_C$ |
|---|---|---|---|---|
| 1 | $\pm 1$ | $\infty_z/m'_z m_x m_{xy}$ | $\infty_z 2'_x 2'_{xy}$ | Time-invariant |
| 1 | 0 or $\pm 2$ | $\infty_z/m'_z m'_x m'_{xy}$ | $\infty_z m'_x m'_{xy}$ | Time-noninvariant |
| 1 | $c \notin \mathbb{Z}$ | $\infty_z/m'_z$ | $\infty_z$ | Time-invariant |
| 2 | $\forall c$ | $m_y m'_x m'_z$ | $2'_x/m'_x$ | No |
| $n \in (-\infty, -1] \cup [3, +\infty)$ | $\forall c$ | $(2|n-1|)'_z/m'_z m'_\xi m_\eta$ | $|n-1|_z m'_\xi m'_\eta$ (odd $n$) | Time-noninvariant |
|  |  |  | $\overline{|n-1|}_z m'_\xi$ (even $n$) | No |

* planes of mirrors $m'_\xi$ and $m'_\eta$ are perpendicular to the $(XY)$ plane.

In order to obtain the symmetry groups $\tilde{G}_V$ and $\tilde{G}_C$ of physical properties and microscopic electronic and spin structure (e.g. the skyrmion at atomic scale [29]), we should intersect groups $G_V$ and $G_C$ with group of the crystal in paramagnetic phase $G_P$ [30]. In case of skyrmions and vortex cores (group $G_C$), symmetry elements $h_i$ changing axial time-odd component $M_z$ must be removed. Set of such operations $\{h_i\}$ includes the inversion centers $\bar{1}'$, rotation axes $r'$ and inversion axes $\bar{r}'$, reflecting mirrors $m'_z$ and reflecting mirrors containing axis $Z$: $m_x$, $m_y$, $m_{xy}$, etc. The $|n-1|$-fold rotation axis appears instead of the $r$.

The antivortices ($n = -1$) appear in the infinity-shaped and clover-shaped samples [34]. Symmetry of the central part of such samples is $G_S = 4/mmm1'$, which affect the magnetization distribution symmetry, according to the theory [41]. Corresponding subgroup, which allows the antivortices symmetry is $4'_z/m'_z m'_\xi m_\eta$, which corresponds (table 1) to the $n = -1$.

### 3. FME coupling in the magnetic vortices and skyrmions

The polarization induced by the FME effect in the cubic m$\bar{3}$m crystal is described by the following four-constant expression [42]:

$$\mathbf{P} = \chi_e[\tilde{\gamma}_1 \mathbf{e}_i \nabla_i(M_i^2) + \tilde{\gamma}_2 \nabla(\mathbf{M}^2) + \tilde{\gamma}_3(\mathbf{M}\nabla)\mathbf{M} + \tilde{\gamma}_4 \mathbf{M}(\nabla\mathbf{M})] \quad (3)$$

Let us consider the vortex and skyrmion structures in plane of the thin film, where exchange energy requires the uniform magnetization distribution in the out-of-plane direction. The in-plane polarization is given by:

$$\mathbf{P}_\perp = \chi_e\{\tilde{\gamma}_1[\mathbf{e_x}\nabla_x(M_x^2) + \mathbf{e_y}\nabla_y(M_y^2)] + \tilde{\gamma}_2\nabla_\perp(\mathbf{M}^2) + \tilde{\gamma}_3(\mathbf{M}_\perp \nabla_\perp)\mathbf{M}_\perp + \tilde{\gamma}_4\mathbf{M}_\perp(\nabla_\perp \mathbf{M}_\perp)\}, \quad (4)$$

where "$\perp$" defines "$x$" or "$y$". The schematic illustration of the induced in-plane polarization was calculated at the $\tilde{\gamma}_3\tilde{\gamma}_4 < 0$ and $\tilde{\gamma}_1^2 \ll |\tilde{\gamma}_3\tilde{\gamma}_4|$ [14,42] (Fig. 2). Vortex ($n = 3$) and antivortex ($n = -1$) have identical magnetic point groups $G_V = 4'/m'_z m'_\perp m_\perp$. They have similar magnetization distributions (Fig. 1e,f). Topologically, their polarization distribution are same, but coupled charges have opposite signs (Fig. 2a,b) [14].

Let us choose the out-of-plane core magnetization [24] in the form:

$$M_z = M_s e^{-\frac{\rho}{\rho_0}}, \quad (5)$$

where $\rho_0$ determines the vortex core size. In scope of the symmetry based analysis, we can define skyrmion as vortex with core size $\rho_0$ of the same magnitude as in-plane size of vortex. In case of the constant magnetization magnitude, it leads to the modification of the expression (1):

$$\mathbf{M}_\perp(\rho, \varphi) = M_s\sqrt{1 - e^{-\frac{2\rho}{\rho_0}}}(\mathbf{e_x}\cos(n\varphi + c\pi/2) + \mathbf{e_y}\sin(n\varphi + c\pi/2)), \qquad (6)$$

The out-of-plane polarization exists inside the vortex core only:

$$P_z = \chi_e\{\tilde{\gamma}_3(\mathbf{M}_\perp \nabla_\perp)\mathbf{M}_z + \tilde{\gamma}_4\mathbf{M}_z(\nabla_\perp \mathbf{M}_\perp)\} \qquad (7)$$

For the above-mentioned magnetization distribution anzatz, it can be expressed in the following form:

$$P_z = \chi_e M_s^2 \frac{e^{-\frac{\rho}{\rho_0}}\cos[(n-1)\varphi + c\pi/2]}{\rho_0\sqrt{1 - e^{-\frac{2\rho}{\rho_0}}}}\left\{\tilde{\gamma}_3(e^{-\frac{2\rho}{\rho_0}} - 1) + \tilde{\gamma}_4\frac{\rho_0}{\rho}e^{-\frac{2\rho}{\rho_0}}\left[\frac{\rho}{\rho_0} + \left(e^{\frac{2\rho}{\rho_0}} - 1\right)n\right]\right\} \qquad (8)$$

The criterion of the $P_z = 0$ is $n = 1$ and $c = \pm 1$, which corresponds to the symmetry group $G_C = \infty_z 2'_x 2'_{xy}$ (Fig. 1c, 2a). Consequently, the expression (8) conforms to the symmetry based prediction:

$$2'_x \cdot P_z = -P_z \Rightarrow P_z = 0 \qquad (9)$$

Total single side surface bound charge of the $P_z$ is determined by the following expression:

$$Q_{SF} = \int_0^{2\pi} d\varphi \int_0^{+\infty} \rho P_z\, d\rho \qquad (10)$$

In contrast to the volume charge $Q_V$ [14], the $Q_{SF}$ is located only in the vortex core. It is given by the following expression ($n = 1$):

$$Q_{SF} = \frac{\chi_e \rho_0 M_s^2 \pi^2 \cos\frac{c\pi}{2}}{4}(\tilde{\gamma}_4 - \tilde{\gamma}_3)(1 + \log 4), \qquad (11)$$

If $n \neq 1$ or $c = \pm 1$ then $Q_{SF} = 0$. Thus, the surface bound charge $Q_{SF}$ equals zero in case of the vortices with winding number $n > 1$ and all antivortices (Fig. 3). The well-known vortex structure with $n = 1$ and $c = \pm 1$ cannot carry total $Q_{SF}$. Oppositely, vortex of the type $n = 1$ and $c = 0$ carries $Q_{SF}$. This vortex has the large magnetic stray fields and usually is not energetically favorable. However, similar types of the skyrmions were predicted for the crystal with the specific symmetry [23]. The spiral type vortex with $n=1$ and $c \notin \mathbb{Z}$ (Fig. 1b) has small stray fields and can carry the total $Q_{SF}$. In case of the thin films or nanodots, these surface charges form the dipole $\boldsymbol{p} = \mathbf{e_z} Q_{SF} h$, where $h$ is the sample width in the $Z$ direction. This dipole is coupled with core magnetization $M_z$. Switching of the $\boldsymbol{p}$ direction is coupled with switching of the $M_z$ or vortex chirality. Detailed features of this coupling need the clarification via the micromagnetic simulation of the type [24]. Vortices and skyrmions with $n \neq 1$ cannot be sensitive to the inhomogeneous electric field.

As it was stated above, there are symmetry based criteria which can forbid the electric dipole properties of magnetic vortex (skyrmion). Symmetry group must be polar to allow $\boldsymbol{p}$. Oppositely, volume electric charge $Q_V$ cannot be forbidden by some symmetry arguments. Symmetry of electric charge is the highest magnetic point group $\infty\infty m1'$. It is allowed by any symmetry medium. Generally, the charge $Q_V$ can have FME and another microscopic nature. For instance, the quantum Hall skyrmions carrying the electrical charge [43].

The discussed theory can be generalized to the weak ferromagnetics with non-collinear magnetic ordering and to the antiferromagnetics using algorithms specified in [44] and [45] respectively. The same FME interactions are possible in case of the antiferromagnetic ordering as it was shown for the multiferroic bismuth ferrite $BiFeO_3$ [16].

The similar calculations of the FME coupling can be performed for any symmetry crystals of cubic, tetragonal and orthorhombic crystal families. Corresponding phenomenological theory was recently reported [18].

## 4. Conclusions

Thus, the electric dipole moment of the vortex is located inside the small vortex core. Switching of the dipole direction makes switching of the core magnetization or vortex chirality and vice versa. The vortices and skyrmions with time-invariant enantiomorphism have two degenerative states: clockwise and counterclockwise state. The vortices and skyrmions with non-unitary winding number do not carry the core electric dipole induced by the flexomagnetoelectric interaction in the hexoctahedral cubic crystal.

**Acknowledgements**

I would like to express my sincere gratitude to Acad. V.G. Bar'yakhtar, who inspired me to the present work. I thank Prof. V.F. Kovalenko, Prof. V. A. L'vov, Dr. S. V. Olszewski, and Dr. O. V. Tychko for their helpful discussion and suggestions. I thank Prof. V. V. Il'chenko for valuable suggestions in the nanotechnology related applications of the present theory. I thank my wife D. M. Tanygina for support with graphical design.

a)

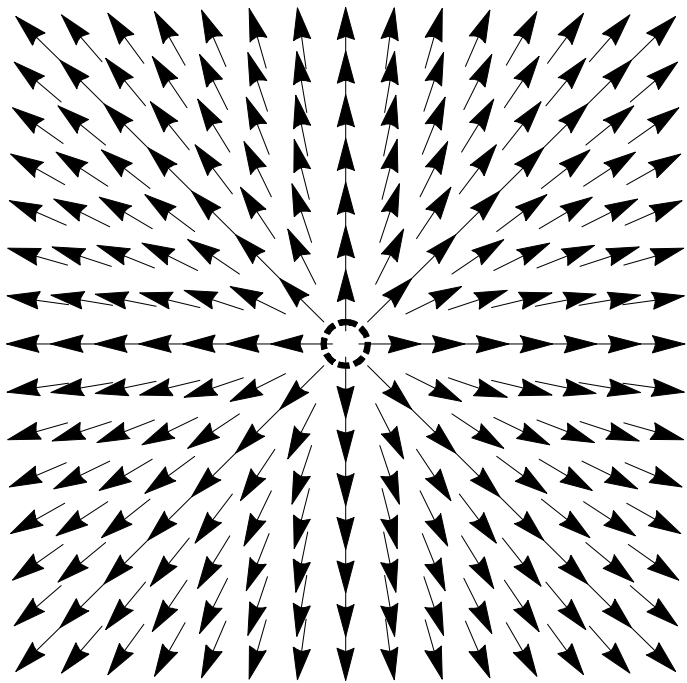

b)

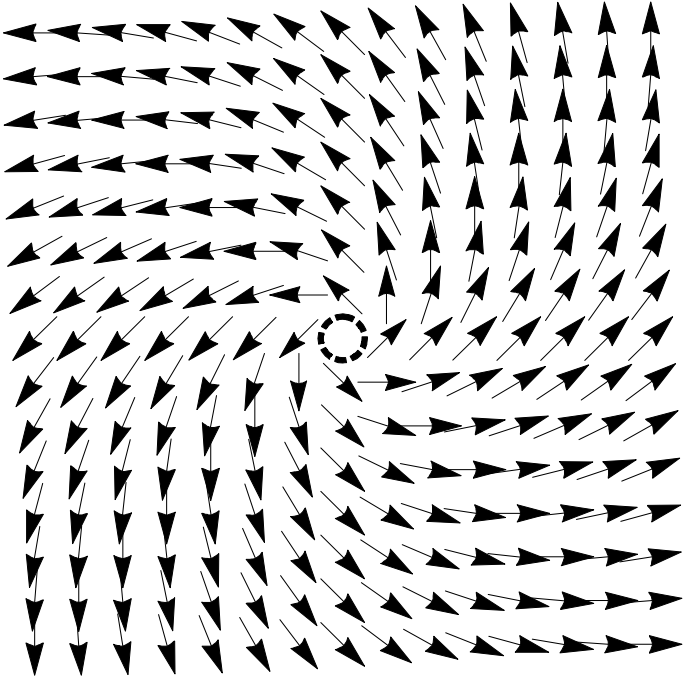

c)

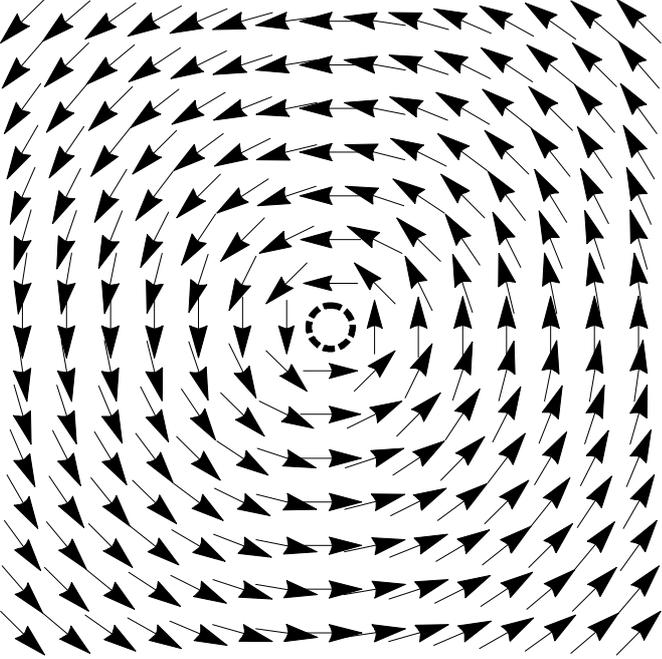

d)

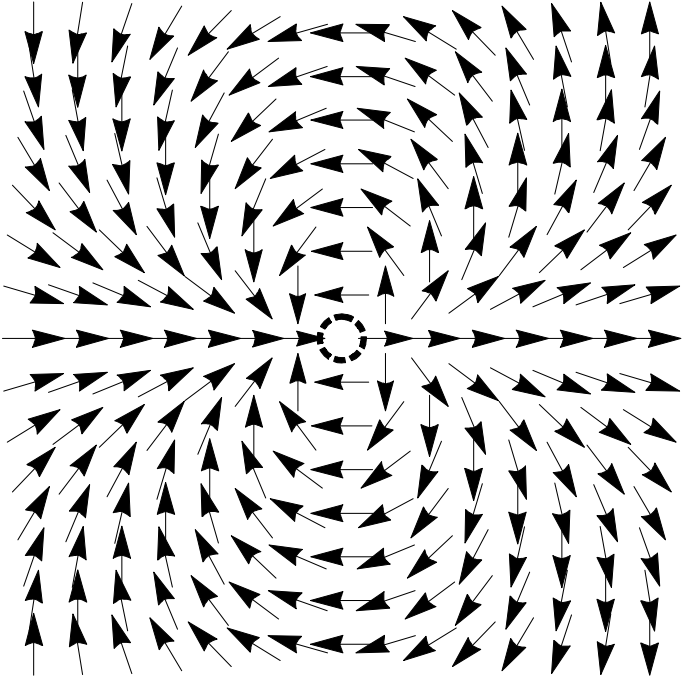

e)

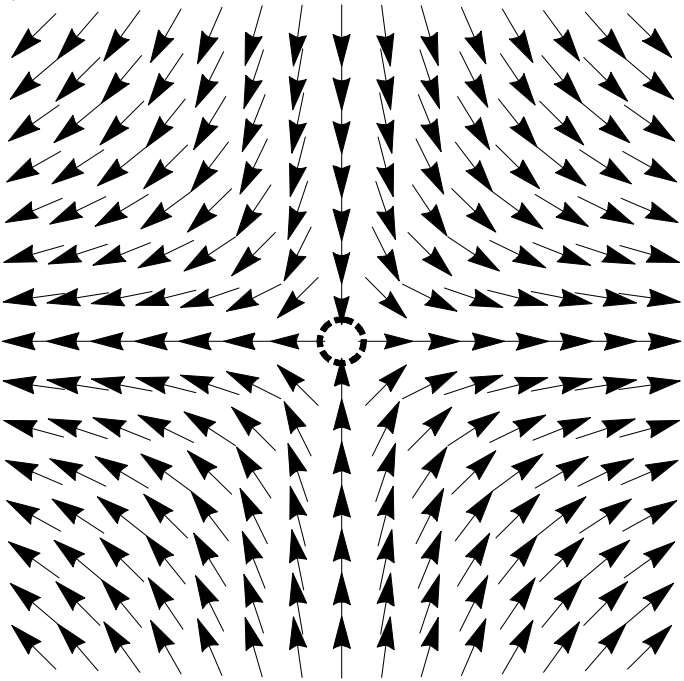

f)

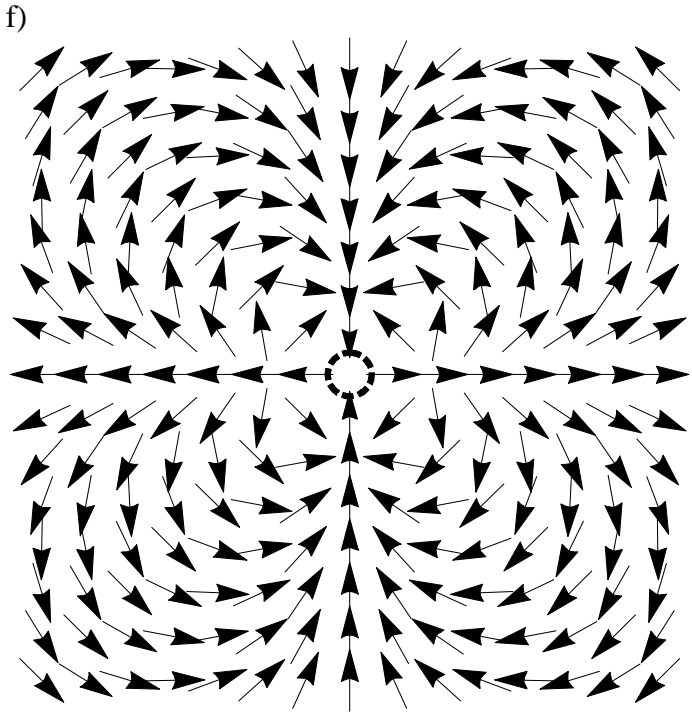

**Fig 1.** Schematic illustration of the in-plane magnetization arrangement in the magnetic vortices (a-d,f) and antivortices (e) with different symmetries around the core (dashed line): $\infty_z/m'_z m'_x m'_{xy}$ (a), $\infty_z/m'_z$ (b), $\infty_z/m'_z m_x m_{xy}$ (c), $m_y m'_x m'_z$ (d) and $4'/m'_z m'_\perp m_\perp$ with winding number $n$ equals -1 (e) and 3 (f).

a)

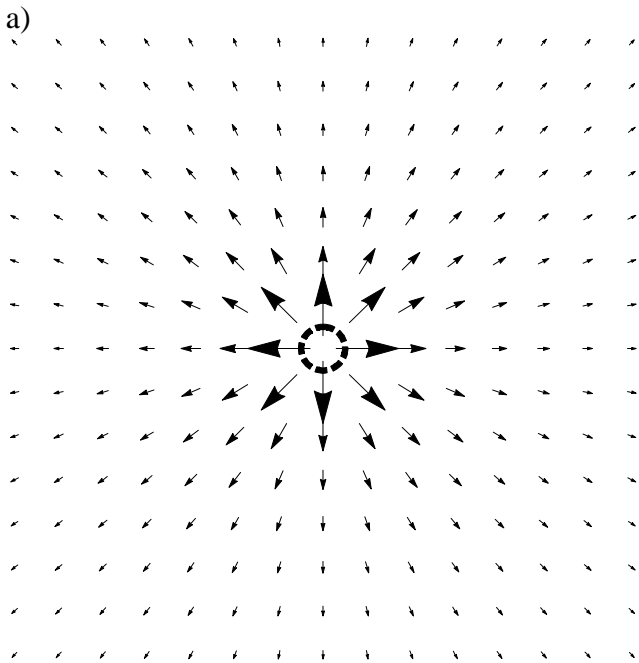

b) 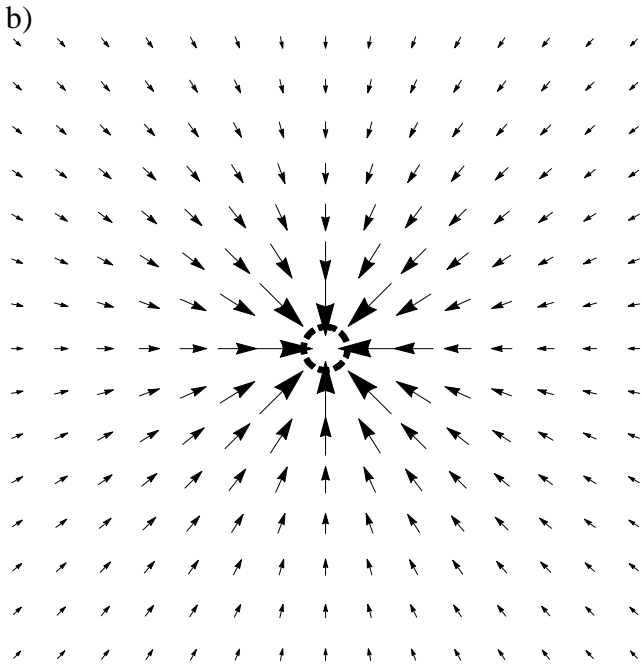

**Fig 2.** Schematic illustration of the in-plane electric polarization in the magnetic vortices (a) and antivortices (b). The winding numbers, respectively, are $n > 0$ (a) and $n < 0$ (b). The core region has been highlighted (dashed line).

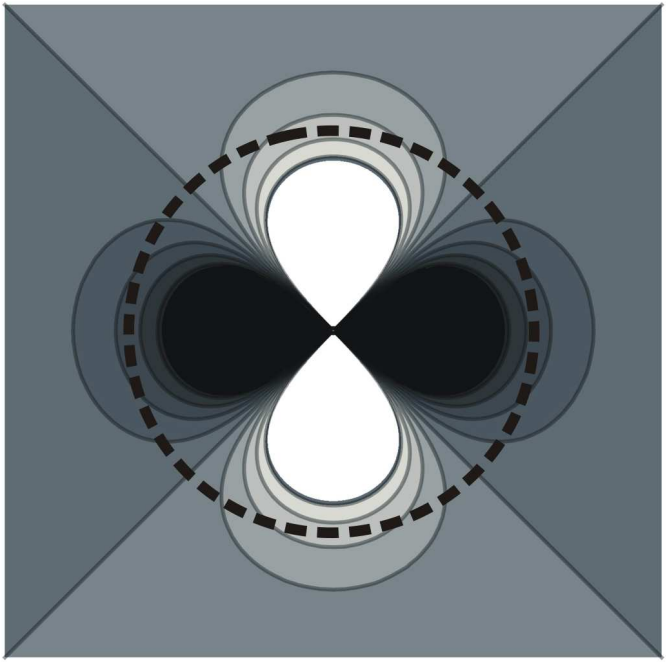

**Fig 3.** Schematic illustration of the out-of-plane polarization in the magnetic antivortex with symmetry $4'/m'_z m'_\perp m_\perp$ at winding number $n = -1$ (shown in gray palette). The black and white correspond to the opposite signs. The core region has been highlighted (dashed line).